\documentclass[12pt]{article}

\usepackage[latin1]{inputenc}
\usepackage{exscale}
\usepackage{amsmath}
\usepackage{amssymb}

\renewcommand{\arraystretch}{1.2}
\newlength{\defbaselineskip}
\newcommand{\qqquad}{\qquad\quad}

\newcommand{\diff}[2][]{\ensuremath{\mathrm{d}^{#1} #2}}
\newcommand{\myint}[2][]{\int\!\diff[#1]{#2} \,}
\newcommand{\ket}[1]{\ensuremath{\left| #1 \right\rangle}}

\newcommand{\braket}[2]{\ensuremath{\left\langle #1 \right|\left.\hspace{-3pt} #2 \right\rangle}}
\newcommand{\eval}[1]{\ensuremath{\left\langle #1 \right\rangle}}
\newcommand{\normord}[1]{\ensuremath{:\!#1\!:}}
\newcommand{\commu}[2]{\left[ #1, #2 \right]}
\newcommand{\anticommu}[2]{\left\{ #1, #2 \right\}}

\newcommand{\vacuum}{\!\ket{0}}

\newcommand{\btheta}{\ensuremath{\bar{\theta}}}
\newcommand{\bxi}{\ensuremath{\bar{\xi}}}
\newcommand{\bLambda}{\ensuremath{\bar{\Lambda}}}
\newcommand{\verylongrightarrow}{-\!\!\!-\!\!\!-\!\!\!-\!\!\!-\!\!\!-\!\!\!-\!\!\!\!\longrightarrow}
\newcommand{\verylongleftarrow}{\longleftarrow\!\!\!\!-\!\!\!-\!\!\!-\!\!\!-\!\!\!-\!\!\!-\!\!\!-}

\newcommand{\unit}{\mathbb{I}}
\newcommand{\tunit}{\tilde{\unit}}
\newcommand{\bcsystem}{$b$-$c$-system}
\newcommand{\bcghostsystem}{$b$-$c$-ghost-system}
\newcommand{\bcghostsystems}{$b$-$c$-ghost-systems}
\newcommand{\ttbsystem}{$\theta$-$\btheta$-system}
\newcommand{\invacuum}{in-vacuum}

\newcommand{\jordanblocks}{Jordan-blocks}
\newcommand{\jordancell}{Jordan-cell}
\newcommand{\jordancells}{Jordan-cells}
\newcommand{\jordanmatrix}{Jordan-matrix}
\newcommand{\jordanrank}{Jordan-rank}
\newcommand{\zeromode}{zero mode}
\newcommand{\zeromodes}{zero modes}
\def\tensor{\otimes}

\begin{document}

\begin{titlepage}
\begin{flushright}
\vspace*{-2cm}{\tt hep-th/0212016}\\
ITP--UH--26/02\\
\end{flushright}

\vspace{1.5cm}

\begin{center}
  {\Large\sc Ghost Systems Revisited:\\[0.5ex]
     Modified Virasoro Generators\\[0.5ex]
     and Logarithmic Conformal Field Theories\\[1.3ex]}
\end{center}

\vspace{1cm}

\begin{center}
  {\sc Marco Krohn, Michael Flohr\footnote{email: krohn@itp.uni-hannvoer.de,
    flohr@itp.uni-hannover.de}}\\[0.5ex]
  {\em Institute for Theoretical Physics\\
       University of Hannover\\
       Appelstraße 2, D-30167 Hannover, Germany}
\end{center}

\vspace{1cm}

\begin{abstract}
We study the possibility of extending ghost systems with higher spin to a
logarithmic conformal field theory. In particular we are interested in 
$c=-26$ which turns out to behave very differently to the 
already known $c=-2$ case. The energy momentum tensor
cannot be built anymore by a combination of derivatives of generalized
symplectic fermion fields. 
Moreover, the logarithmically extended theory is only consistent
when considered on nontrivial Riemann surfaces.
This results in a LCFT with some unexpected properties.
For instance the Virasoro mode $L_0$ is diagonal and for certain values
of the deformation parameters even the whole global conformal group is non-logarithmic.
\end{abstract}
\end{titlepage}

\newpage


\section{Introduction}
Only three years after Belavin, Polyakov and Zamolodchikov \cite{Belavin:1984vu} 
started investigating conformal field theories
in two dimensions it was noted by Knizhnik \cite{Knizhnik:1987xp} that correlation functions may
also exhibit logarithmic divergences. Six years later Gurarie \cite{Gurarie:1993xq} introduced
the concept of a conformal field theory with logarithmic singularities: operator product
expansions have the general form
\begin{align}
  A(z)B(0) = z^{h_C-h_B-h_A} \sum_n \log^n (x) \left\{ C_{m-n} + \ldots \right\}
\end{align}
where $C_n$ denotes the so-called logarithmic partners 
for $0 \le n < m$ and $m$ gives the size
of the \jordancell, i.\,e., the number of logarithmic partner fields which
together with the primary field span a \jordancell\ for the \zeromode\ $L_0$ of the
Virasoro algebra. Thus, 
the basic feature of these so-called logarithmic conformal
field theories (LCFTs) is that the representations of the chiral symmetry
algebra may be indecomposable. It is worth noting that various aspects
of LCFTs were noted in the literature before the work of
Gurarie, e.\,g.\ \cite{Rozansky:1993td,Saleur:1992hk,Saleur:1992vh}.

Logarithmic conformal field theories have a number of applications in very
different fields, such as fractional quantum Hall effect,
gravitational dressing, disorder, string theory and AdS/CFT
to name only a few. For some recent reviews see \cite{Flohr:2001zs,
Gaberdiel:1999mc,Moghimi-Araghi:2002gk,Kawai:2002fu,RahimiTabar:2001} and references therein.

Our own interest is related to the well-known \bcghostsystem\ 
which is a quantum field theory of two anti-commuting fields $b,c$ with integer
or half-integer spins $\lambda$ and $1-\lambda$, respectively. The bosonic
$\beta$-$\gamma$-ghost-systems have recently been studied by F.~Lesage, P.~Mathieu {\em et al.}\
\cite{Lesage:2002ch}.
It is conventional to call the spin $\lambda$ field $b(z)$
and the field with spin $1-\lambda$ then $c(z)$ with the choice $\lambda \ge 1-\lambda$.
In this paper, we are interested in the integer spin case only.
Of course correlators of the $b$ and $c$ fields do not show logarithmic
behavior, but we claim that it is possible to enlarge the system
to a logarithmic conformal field theory. To be more precise: we
conjecture that each \bcsystem\ (with central charge
$c_{b,c} = 2(-1 + 6\lambda - 6\lambda^2)$) is in fact a subset of a
larger logarithmic CFT with \jordancells\ of higher rank related to the spin
$\lambda>0$. We will in detail study the
case $c=-26$ which is the next integer spin case, $(\lambda,1-\lambda)=(2,-1)$,
after the well-known $c=-2$ theory with spin $(\lambda,1-\lambda) = (1,0)$.
Indeed we find a nontrivial indecomposable structure
of the Virasoro modules which, however, is quite
different from the \jordancell\ structure
in the $c=-2$ system. 

Our paper will proceed as follows:

In the next section we will briefly recall the main properties 
of the $c=-2$ LCFT. First we have a look at the construction
via symplectic fermions. We then review an alternative approach 
\cite{Fjelstad:2002ei} where 
the $c=-2$ LCFT is built via deformation of the energy momentum tensor.

The third section then deals with the generalization of these procedures to 
the ghost system with
central charge $c=-26$. Firstly, we consider the \zeromode\
structure of the fields by using a generalization 
of the symplectic fermion method. It turns out that the energy-momentum
tensor cannot be constructed in a similar fashion as in the $c=-2$ case 
out of these fields, without running into severe difficulties.
However, a generalization of the method of deformation is possible and yields a 
consistent representation of the Virasoro algebra. 
Thus, the two approaches are not equivalent in the
$c=-26$ case. 

Unfortunately, this Virasoro algebra does not act consistently
on the Hilbert space of states of this theory. The reason for this is related
to the origin of the logarithmic operators, which arise from operator
product expansions of twist fields \cite{Kogan:1998fd}. These twist fields
exist whenever the theory is put on a nontrivial Riemann surface
\cite{Knizhnik:1987xp}. Thus,
we investigate the theory on the simplest nontrivial Riemann 
surfaces, the hyper-elliptic ones, and find that the full theory 
features a consistent Virasoro algebra with the correct action on its
space of states. Although this full theory turns out to be logarithmic, its
structure is very different from the $c=-2$ case. For example, the \zeromode\ 
of the Virasoro algebra, $L_0$, turns out to be diagonal, i.\,e., the
Virasoro modules are \emph{not\/} indecomposable with respect to $L_0$. 
However, other Virasoro modes definitely lead to indecomposable structures.
The section concludes with building highest weight states for
different conformal weights and discussing a suitable generalization of
the \jordanrank\ of the theory.


\section{The {\boldmath $b$, $c$ ghost system as subset of logarithmic $c=-2$} theory}
The conformal \bcsystem\ and the associated logarithmic so called \ttbsystem\ 
for central charge $c=-2$ are well-known and have been intensely studied 
(see e.\,g.\ \cite{Cappelli:1998ma,Flohr:2000mc,Gurarie:1993xq,
Gurarie:1997dw,Kausch:1991bn,Kausch:1995py}).
This is the reason for us having a closer look at this system again
in the hope of learning how to build such logarithmic theories in general.
In the case of $c=-2$ we will briefly repeat two different ways of building
a LCFT: firstly via symplectic fermions \cite{Gurarie:1997dw,Kausch:2000fu},
and secondly by deforming the energy momentum tensor \cite{Fjelstad:2002ei}.

\subsection{{\boldmath $c=-2$} LCFT via Symplectic Fermions}
  
Following the approach described in \cite{Gurarie:1997dw} the $c=-2$ theory can be
represented as a pair of ghost fields, or anti-commuting fields $\theta$,
$\btheta$ of conformal weight $h=0$, with the free action \cite{Gurarie:1993xq}
\begin{align}
  S = \myint[2]{z} \partial\theta \bar{\partial}\btheta \quad .
\end{align}
(Note that $\theta$, $\btheta$ are \emph{not} the complex conjugate
of each other, but different fields.) As described in the  
above mentioned reference, the vacuum \ket{0} is somewhat unusual, 
its norm is $\braket{0}{0} = 0$, while the explicit insertion of the 
fields $\theta$ produces nonzero results, for instance
$\eval{\btheta (z){\theta (w)}} = 1$. This property of the vacuum is
believed to be typical for LCFTs.

Using the results given in \cite{Gurarie:1997dw} the mode expansion of the
field $\theta$ (the analog holds for $\btheta$) is
\begin{align}
  \theta (z) = \xi + \theta_0 \log (z) + \sum_{n \ne 0} \theta_n z^{-n} \quad ,
\end{align}
where $\xi$ denotes the crucial \zeromodes\ and $n \in \mathbb{Z}$.
The non-vanishing anti-commutators ($n \in \mathbb{Z}$, $n\ne 0$) are
\begin{align}
  \anticommu{\theta_n}{\btheta_m} = \frac 1n \delta_{n,-m} \, , \quad
  \anticommu{\xi}{\btheta_0} = 1 \, , \quad
  \anticommu{\theta_0}{\bxi} = -1 \, ,
\end{align}
and together with the highest-weight relation
\begin{align}
  \theta_n \ket{0} = 0 \quad \forall n \ge 0 \quad ,
\end{align}
it is quite easy to see the logarithmic nature of the $\theta$, $\btheta$
system, for instance by calculating
\begin{align}
  \eval{\tunit (z) \tunit (w)} & = -2 \log (z-w)
\end{align}
where $\tunit$ is defined as $\tunit \equiv - \normord{\theta\btheta}$ .

The stress energy tensor of the theory is
\begin{align}
  T(z) = \normord{\partial\theta \partial\btheta} \label{eq:T(z)_for_theta_btheta}
\end{align}
and it is not hard to see that its expansion with $\tunit$ is indeed
given by 
\begin{align}
  T(z)\tunit(w) = \frac{\unit}{(z-w)^2} + \frac{\partial\tunit(w)}{z-w} + \ldots
\end{align}
meaning that the operator $\tunit$ has conformal weight 0.
Also $\tunit$ is the logarithmic partner of $\unit$, since
$L_0 \tunit = \unit$. Thus, $\unit$ and $\tunit$ span a \jordancell\ of
rank two with respect to $L_0$. Indeed, the reader should convince herself
that the action of $L_0$ cannot be diagonalized.

The most obvious differences between the \bcsystem\ and the \ttbsystem\ 
are 

\begin{center}
\begin{tabular}{lcc}
  \zeromodes: & ($b_0$, $c_0$) & ($\btheta_0$, $\xi$), ($\theta_0$, $\bxi$) \\
  conformal weights: & $h(b)=1$, & $h(\btheta)=0$,  \\
                    & $h(c)=0$ & $h(\theta)=0$ 
\end{tabular}
\end{center}

Therefore, in order to get from the \ttbsystem\ to the \bcsystem\ 
we have to reduce the number of \zeromode\ pairs by one and also 
have to increase the conformal weight of one of the fields by one.
This can easily be done by defining the transformation between
$b$, $c$ and $\theta$, $\btheta$ in the following way:
\renewcommand{\arraystretch}{0.5}
\begin{eqnarray}
  b(z)=\sum_{n\in\mathbb{Z}} b_n z^{-n-1} &
    \begin{array}{c} 
      \scriptstyle{\btheta = \partial^{-1}b} \\
      \verylongrightarrow \\ 
      \verylongleftarrow  \\    
      \scriptstyle{b = \partial \btheta}
    \end{array} &
    \btheta(z)=\sum_{n\ne 0} \btheta_n z^{-n} + \btheta_0 \log(z) + \bxi 
    \label{eq:bc_lcft_map_b} \\ 
  \nonumber \\
  c(z)=\sum_{n\in\mathbb{Z}} c_n z^{-n} &
    \begin{array}{c} 
      \scriptstyle{\theta = c + \theta_0\log(z)} \\
      \verylongrightarrow \\ 
      \verylongleftarrow  \\    
      \scriptstyle{c=\theta|_{\theta_0=0}}
    \end{array} &
    \theta(z)=\sum_{n\ne 0} \theta_n z^{-n} + \theta_0 \log(z) + \xi \; .
    \label{eq:bc_lcft_map_c}
\end{eqnarray}
While the derivative (respectively integration) gives the right transformation
between $b$ and $\btheta$ we artificially have to add (respectively eliminate) a \zeromode, 
$\theta_0$, to get the transformation between $c$ and $\theta$.

One might be tempted to use this method for constructing higher 
logarithmic CFTs, namely by putting the $b$, $c$ fields on equal footing
by integrating the $b$ field $2 \lambda + 1$ times where $\lambda >0$ denotes
the conformal weight of the $b$ field. This 
integration leaves us with $2 \lambda + 1$  new modes which then turn out to be 
one half of the total set of \zeromodes. The other half of the \zeromodes\ 
has to be added artificially in an analogous way, as for the $c$ field
shown above. The latter are necessary as canonically conjugate partners for 
the \zeromodes\ arising as integration constants. Without these conjugate 
partners, the action of our \zeromodes\ would be trivial.

\subsection{{\boldmath $c=-2$} LCFT via logarithmic deformation}
As noted in the introduction of this section there is a 
different way to construct logarithmic extensions of conformal field
theories as described in
\cite{Fjelstad:2002ei}. The idea of this method is to consider
special deformations of the energy momentum tensor. One
defines
\begin{align}
  \tilde{T} := T^{\mathrm{CFT}} + T^{\mathrm{impr}}
\end{align}
where $T^{\mathrm{impr}}$ denotes the so called
``improvement term'' which extends the CFT energy momentum
tensor $T^{\mathrm{CFT}}$ in a way that the resulting
stress tensor $\tilde{T}$ belongs to a logarithmic theory.
Of course, the full stress energy tensor must still possess the
correct operator product expansion with itself.

As is well-known the CFT stress tensor is given by
\begin{align}
  T^{\mathrm{CFT}} = -\lambda\normord{b(\partial c)} + (1-\lambda)\normord{(\partial b) c} \,\,
                   = -\normord{b\partial c} \quad ,
\end{align}
for $\lambda=1$, which yields the $c=-2$ ghost-system.
A careful consideration motivates the following ansatz
for the improvement term:
\begin{align}
  T^{\mathrm{impr}} = \frac 1z \theta_0 b(z)
\end{align}
with $\theta_0$ being an additional \zeromode. We have deliberately
chosen to name this \zeromode\ $\theta_0$ to make contact to the preceding
approach via symplectic fermions. Indeed, the deformed
energy momentum tensor can in this case be rewritten in a nicer form
by applying a deformation to the fields as well:
\begin{eqnarray}
  b(z) & \longrightarrow & \btheta(z) = \partial^{-1} b(z) \label{eq:redef_b}\\
  c(z) & \longrightarrow & \theta(z) = c(z) + \theta_0 \log(z)\label{eq:redef_c}
\end{eqnarray}
which leads to the well-known result (\ref{eq:T(z)_for_theta_btheta}).
The theory with $c=-2$ is a bit special, because (as we will see later) it is 
not always possible to write the energy momentum tensor as a 
function of the new basic fields.


\section{A close look at {\boldmath $c=-26$}}
Motivated by the success for the $c=-2$ system we are now going to
construct a logarithmic conformal field theory for $c=-26$ which
basically has the same properties as the $\theta$, $\btheta$ system
in the $c=-2$ case. This construction process presumably does not only work
for $c=-26$, but should work for any \bcghostsystem .

One might be tempted to assume that the \jordancell\ of the LCFT for $c=-26$ has
a rank greater than two since this theory possesses a larger number of
\zeromodes, i.e., modes which annihilate the vacuum to the left as well as to the right.

As we will see we do not find
higher rank \jordanblocks: in fact, the \zeromode\ of 
the Virasoro algebra $L_0$ turns out to be perfectly 
well-defined without any Jordan structure at all.
The nontrivial indecomposable structure of the Virasoro
modules manifests itself in the action of the Virasoro modes
$L_n, n \ne 0$. Therefore, we cannot speak of
a rank of a \jordancell\ anymore. We will discuss later in which way the Virasoro modules
are indeed indecomposable. 

Investigating such systems is interesting for string theory.
The calculation of string amplitudes makes use of the computation of
$\lambda$-forms on nontrivial Riemann-surfaces. In a CFT approach
these are the ghost systems. As Knizhnik \cite{Knizhnik:1987xp} pointed out,
nontrivial
Riemann-surfaces, seen as a multisheeted covering of the complex
plane, can be simulated by twist-fields inserted at
the branch points. It has become clear by now that operator product
expansions of such twist fields inevitable lead to logarithmic fields \cite{Gaberdiel:1996np, Flohr:1998ew}.
Therefore, computation of string amplitudes automatically
involves not only the $b$,$c$ system but its enlarged full LCFT.
Also, there have been hints that
LCFTs with higher rank \jordanblocks\ play a role in the AdS/CFT
correspondence \cite{Giribet:2001qq}. Thus, it is important to learn more
about LCFTs where the indecomposable structure is more involved than in the
simple rank-two case. Even the simplest such higher-rank cases are very
difficult to study, since the generic form of operator product expansion 
can only be fixed under quite restrictive assumptions \cite{Flohr:2001tj}.

\subsection{Generalizing symplectic fermions}
We now try to mimic what we did in the previous section, but 
this time for $c=-26$. Starting with the
well-known (e.\,g.\ \cite{Ket95}) \bcsystem\ for $c = -26$ and by applying the
same steps as we did for $c=-2$ we get a larger system. Unfortunately building
a LCFT for $c=-26$ turns out to be more complicated than for the $c=-2$ case.
Basically two obstacles are in the way of constructing a LCFT for $\lambda \ge 2$:

\begin{enumerate}
  \item The energy momentum tensor cannot be built by combining derivatives
     of the generalized symplectic fermion fields.
  \item LCFT is intimately linked to twist fields arising from putting the
     CFT on a nontrivial Riemann-surfaces. The full theory is a tensor product
     of the CFTs for each covering sheet. We cannot neglect this fact. 
\end{enumerate}

In this sense $c=-2$ is special since the above mentioned problems do
not show up (as we will explain later).

The \bcsystem\ for $c=-26$ is given by the following relations
if we set $\lambda = 2$:
\begin{eqnarray}
  b(z) = \sum_n b_n z^{-n-\lambda}\,, && c(z)=\sum_n c_n z^{-n-(1-\lambda)} \label{eq:bc-field} \\
  \anticommu{c_n}{b_m} = \delta_{n+m,0}\, , && \anticommu{c_n}{c_m} = \anticommu{b_n}{b_m} = 0 \\ 
  \label{eq:bc-hw-cond}
  b_n \ket{0} = 0 \quad\forall n\ge -\lambda+1\,, && c_n \ket{0} = 0 \quad \forall n \ge \lambda\\
  T(z) = -\lambda \normord{b\partial c} \!\!\!&+&\!\!\! (1-\lambda)\normord{(\partial b) c}
    \label{eq:T(z)_for_b_c_c=-2} \quad .
\end{eqnarray}
Note that the $c=-26$ \bcsystem\ comes with three pairs of 
\zeromodes,\footnote{If not explicitly stated otherwise 
the range of $i$ for the $c=-26$ system is -1, 0, 1} namely ($b_i$, $c_i$) for
$i=-1,0,1$. These modes are called \zeromodes\ for the following reason.
As we can infer from the highest-weight conditions (\ref{eq:bc-hw-cond}),
the $b_i$ modes are annihilators to the left and to the right, while the
$c_i$ modes are creators to the left and to the right. Thus, the $b_i$ are
proper \zeromodes, and the $c_{-i}$ are their canonically conjugate partners.
In the same way as we extended the $c=-2$ theory to a larger one by formal 
integration, we can try this for the $c=-26$ case by introducing the fields
\begin{eqnarray}
  \Lambda(z) & := & \theta_1 \log z + \theta_0 ( z \log z - z ) + \theta_{-1} \frac{z²}{2} \left( \log z - \frac 32 \right) \nonumber \\
  &&  + \xi_1 + \xi_0 z + \xi_{-1}\frac 12 z^2 + \sum_{|n|>1} \theta_n \frac{z^{-n+1}}{-n+1} \\
    \bLambda(z) & := & \btheta_1 \log z + \btheta_0 ( z \log z - z ) + \btheta_{-1} \frac{z²}{2} \left( \log z - \frac 32 \right) \nonumber \\
  &&  + \bxi_1 + \bxi_0 z + \bxi_{-1}\frac 12 z^2 + \sum_{|n|>1} \btheta_n \frac{z^{-n+1}}{-n+1} \quad .
\end{eqnarray}
The field $\bLambda(z) := \partial^{-3} b$ has now the same conformal weight 
as its partner field $\Lambda(z) := c(z) + \sum_i f_i(z) \theta_i$. We call 
such pairs of anti-commuting fields of identical conformal weight 
generalized symplectic fermions. Note that the threefold-integration
adds three new modes, $\xi_i$, $i=-1,0,1$, to the theory,
which are (as we will see later) one half of the additional \zeromodes\ we 
have to add to the theory in order to make it logarithmic. The other half is
artificially added in the $\Lambda$ field. Similar to the $c=-2$ case our new
fields are now on equal footing $h(\Lambda) = h(\bLambda) = -1$.

Going from $\Lambda$, $\bLambda$ back to $b$, $c$ of course requires 
removing these additional modes:
\begin{eqnarray}
  b(z) & = & \partial^3 \bLambda(z) \\
  c(z) & = & \Lambda \big|_{\theta_i=0} \quad .
\end{eqnarray}
The relations from the \bcsystem\ can be translated to the new system and
we find:
\begin{eqnarray}
  \anticommu{\theta_n}{\btheta_m} &=& - \frac 1n \delta_{n, -m} \qquad |n|, |m| > 1 \\
  \anticommu{\xi_{i}}{\btheta_{-i}} &=& (-1)^{i+1} \qqquad i = -1,0,1 \quad .
\end{eqnarray}
For the new modes we require the anti-commutation relations to be
\begin{align}
  \anticommu{\bxi_{i}}{\theta_{-i}} := (-1)^i \qqquad i = -1,0,1 \quad ,
\end{align}
which leads to the following OPEs
\begin{eqnarray}
  \Lambda(z) \bLambda(w) & \sim & \frac 12 (z-w)^2 \left[ \frac 34 - \log(z-w) \right] \\
  \Lambda(z) \Lambda(w)  & \sim & \mathcal{O}(z-w) \\
  \bLambda(z) \bLambda(w)  & \sim & \mathcal{O}(z-w) \quad .
\end{eqnarray}
The new modes indeed have the properties of \zeromodes, namely that all
modes $\theta_i$ and $\bar\theta_i$ are annihilators to both sides, and
the modes $\bar\xi_{-i}$ and  $\xi_{-i}$ are their respective conjugate
modes. Therefore,
the extended theory also contains twice as many \zeromodes\ compared
to the original \bcghostsystem.

\subsection{Building the energy momentum tensor}
Having constructed fields which show logarithmic behavior leads
us to the question how the energy momentum tensor for $c=-26$
looks like. Therefore, we look back to the $c=-2$ case
in the hope to learn from this scenario. We remember that
for $\lambda=1$ respectively $c=-2$ simply plugging in the fields
(\ref{eq:bc_lcft_map_b}), (\ref{eq:bc_lcft_map_c}) in the energy momentum
tensor
\begin{align}
  T = T[b,c] = -\lambda \normord{b \partial c} + (1-\lambda) \normord{(\partial b)c} \quad .
\end{align}
gives us the
desired result (\ref{eq:T(z)_for_theta_btheta}). Unfortunately this does
not work out in the same way for $c=-26$ and presumably neither
for any other $\lambda \ge 2$.

The reason is obvious: $\Lambda$ appears plainly and as first derivative
in the energy momentum tensor. Because of $\Lambda$ containing $z^n \log(z)$ terms
this inevitable leads to logarithmic terms in the energy momentum tensor.

To find possible energy momentum tensors at all we use a different approach
and consider possible extensions of the stress tensor on the mode level. 
This approach is motivated by the paper of Fjelstad, 
Fuchs \emph{et al.}\cite{Fjelstad:2002ei}, but note that our 
\emph{deformation term} is slightly more general and so is our result. The
deformation term in \cite{Fjelstad:2002ei} is always constructed from
primary fields, which we do not assume here.
\begin{eqnarray}
  T^{\mathrm{log}}(z) & = & T^{\mathrm{bc}}(z) + R(z) \\
                     & = & \sum_{n} z^{-n-2} \left( L_n^{\textrm{bc}} + R_n \right)
\end{eqnarray}
where the modes $L_n^{\textrm{bc}}$ are given by
\begin{eqnarray}
  L_{-2}^{\textrm{bc}} & = & -\sum_{l \ne -3,\ldots ,1} \frac{l(l+1)(l+4)}{l+3} \normord{\btheta_l \theta_{-l-2}}
       - 6 \normord{\btheta_{-3}\xi_{1}} -4 \normord{\btheta_{-2}\xi_{0}} \nonumber \\
  & &  - \frac 32 \normord{\btheta_{-1}\xi_{-1}} + \frac 43 \normord{\btheta_{0}\theta_{-2}}
       - \frac 52 \normord{\btheta_{1}\theta_{-3}} \\
  L_{-1}^{\textrm{bc}} & = & - \sum_{l \ne -2,\ldots ,1} l(l+1) \normord{\btheta_{l}\theta_{-l-1}}
       - \normord{\btheta_{-1}\xi_{0}} +\normord{\btheta_{0}\xi_{-1}} - 2\normord{\btheta_{1}\theta_{-2}} \\
  L_{0}^{\textrm{bc}} & = & - \sum_{l \ne -1,\ldots ,1} l^2 \normord{\btheta_{l}\theta_{-l}}
       - \normord{\btheta_{1}\xi_{-1}} + \normord{\btheta_{-1}\xi_{1}} 
      \\
  L_{1}^{\textrm{bc}} & = & - \sum_{l \ne -1,\ldots ,2} (l-2)(l+1) \normord{\btheta_{l}\theta_{-l+1}}
       - 3 \normord{\btheta_{-1}\theta_{2}} -2 \normord{\btheta_{0}\xi_{1}}
       - 2 \normord{\btheta_{1}\xi_{0}} \\
  L_{2}^{\textrm{bc}} & = & - \sum_{l \ne -1,\ldots ,3} \frac{(l-4)l(l+1)}{l-1}
                            \normord{\btheta_{l}\theta_{-l+2}}
       - \frac 52 \normord{\btheta_{-1}\theta_{3}} + 4 \normord{\btheta_{0}\theta_{2}} \nonumber \\
  & &  + 6 \normord{\btheta_{1}\xi_{1}} +12 \normord{\btheta_{2}\xi_{0}} 
       + 6 \normord{\btheta_{3}\xi_{-1}}
\end{eqnarray}
and $R_n$ denotes the extension which may contain the new 
\emph{deformation modes} $\btheta_i$, $\xi_i$. 
The modes $L^{\mathrm{log}}_n := L_n + R_n$ of course 
have to obey the Virasoro Algebra, which is a strong restriction.
We get two different solutions, each coming with three possible
deformations of the stress tensor:
\begin{eqnarray}
  R_{-2} &=& 6A \theta_1\btheta_{-3} - 4B \theta_0\btheta_{-2} + \frac 32 C \theta_{-1}\btheta_{-1}  \\
  R_{-1} &=& -B \theta_0\btheta_{-1}  - C \theta_{-1}\btheta_0 \\
  R_{0} &=& -A \theta_1\btheta_{-1}  + C \theta_{-1}\btheta_{1} \\
  R_{1} &=& 2A\theta_{1}\btheta_{0} + 2 B \theta_{0}\btheta_{1} \\
  R_{2} &=& -6A\theta_{1}\btheta_{1}   + 12 B\theta_{0}\btheta_{2} - 6C \theta_{-1}\btheta_{3} 
\end{eqnarray}
\begin{eqnarray}
  R_{-2} &=& 6A' \theta_1\btheta_{-3} - B' \theta_{-2}\bxi_0 
          + \frac 32 C' \theta_{-1}\btheta_{-1} \label{eq:second_sol_start}\\
  R_{-1} &=& - C' \theta_{-1}\btheta_0 \\
  R_{0} &=& -A' \theta_1\btheta_{-1}  + C' \theta_{-1}\btheta_{1} \\
  R_{1} &=& 2A'\theta_{1}\btheta_{0} \\
  R_{2} &=& -6A'\theta_{1}\btheta_{1} -3B' \theta_2 \bxi_{0} 
            - 6C' \theta_{-1}\btheta_{3} \quad .\label{eq:second_sol_end}
\end{eqnarray}
Testing the Virasoro Algebra with the above \emph{deformation terms} is
sufficient, since all higher modes can be derived with the help
of the Virasoro Algebra.

Two things are noteworthy: firstly, the second solution contains $\xi$ modes.
This is a bit unexpected since $L^{\mathrm{log}}_n$ should, according to
what we learned from the $c=-2$ theory, only lower the \zeromode\ content and not
increase it. Secondly, both solutions look very similar. 
Setting $B'=0$ in the second solution, and thus eliminating the unwanted
$\xi$-modes, would result in a special case ($B=0$) of the first solution. 
As we will see later the second solution is indeed a special case of the 
first one. That is why we concentrate on the first solution for now.

The extensions can be written in a nicer way, making 
use of the $b$-field:
\begin{align}
  T^{\mathrm{log}}(z) = T^{\mathrm{bc}}(z) + A \theta_1 \frac{1}{z^0} \partial(z^0 b)
    + B\theta_0 \frac{1}{z^1} \partial(z^2 b) + C\theta_{-1} \frac{1}{z^2} \partial(z^4 b)
    \label{eq:Tlog_b_theta}
\end{align}
which has a strikingly similarity with the energy momentum tensor
deformations described by Fjelstad, Fuchs \emph{et al.}, but also has
an important difference, namely the appearance of derivatives of the first order.
The important point is that the deformations involve additional modes
which are proper \zeromodes, i.\,e.\ annihilation operators to both sides.
There are three possible
``directions'' to deform the energy momentum tensor, which matches exactly
the number of \zeromodes\ of our system as we might have expected. 
In the $c=-2$ system only one such
deformation was possible. There is another difference between $c=-2$
and $c=-26$: while in the former theory it was possible to redefine the $b$
and $c$ fields
(\ref{eq:redef_b}), (\ref{eq:redef_c}) in order to get an energy momentum 
tensor which consists of the new fields only, this is not possible in the
latter case.

Demanding that the Virasoro modes satisfy the 
hermiticity condition $L^\dagger_n = L_{-n}$ leads 
to a further restriction of 
the solution\footnote{Note that taking the adjoint of the modes can
cause an additional constant, for instance $\xi_1^\dagger = \frac 12 \xi_{-1}$,
due to our normalization of the modes, which results
from viewing them as integration constants.}:
\begin{align}
  A = C \quad .
\end{align}
In the second solution, this requirement leads to the condition $A' = C'$.

\subsection{Fields on nontrivial Riemann Surfaces}
Up to now we have constructed fields $\Lambda$, $\bLambda$ out of 
the \bcsystem\ for $c=-26$ and we have found possible deformations of the 
energy-momentum tensor. The Hilbert-space $\mathcal{H}^{\mathrm{log}}$ 
of the extended theory is an enlargement of the Hilbert-space 
of the \bcsystem\ containing the additional \zeromodes\ $\bxi_i$.

This gives rise to another problem, namely that the constructed 
theory cannot be the full theory, because of $L_0^{\textrm{log}}$ 
not being able to measure the conformal weight of all 
states contained in the Hilbert-space correctly. 
For instance $\ket{\bxi_{-1}}$ is surely an element 
of the Hilbert-space $\mathcal{H}^{\mathrm{log}}$,
but $L_0^{\textrm{log}} \ket{\bxi_{-1}} = 0$ gives the wrong conformal weight.

This is an extremely interesting observation. The origin of logarithmic
fields is tied to the existence of so-called pre-logarithmic primary fields,
whose operator product expansions contain the logarithmic fields 
\cite{Kogan:1998fd}. In fact, the first hint for the existence of the
field $\tunit$ in the $c=-2$ theory comes from evaluating the four-point
function of four $\mathbb{Z}_2$ twist fields $\mu$ of conformal weight
$h=-1/8$, as has been observed in \cite{Gurarie:1993xq}. As a result, this
four-point function contains the following two conformal blocks:
\begin{eqnarray}\label{eq:twist4pt}
& &\langle\mu(\infty)\mu(1)\mu(x)\mu(0)\rangle\ =\
   [x(1-x)]^{\frac14}F(x)\,,\nonumber\\
& &F(x) = \left\{\begin{array}{lcl}
   {}_2F_1(\tfrac{1}{2},\tfrac{1}{2};1;x)\,, & &\\[6pt]
   {}_2F_1(\tfrac{1}{2},\tfrac{1}{2};1;1-x) &=&
   {}_2F_1(\tfrac{1}{2},\tfrac{1}{2};1;x)\log(x)\\[6pt]
&+& \left.\!\frac{\partial}{\partial\epsilon}\,
   {}_3F_2(\tfrac{1}{2}\!+\!\epsilon,\tfrac{1}{2}\!+\!\epsilon,1;1\!+\!\epsilon,
   1\!+\!\epsilon;x)\right|_{\epsilon=0}\,.\end{array}\right. 
\end{eqnarray}
In case of the
ghost systems, these pre-logarithmic twist fields have a geometric meaning:
these fields behave exactly as branch points of a ramified covering of
the complex plane. For example, the above mentioned $\mathbb{Z}_2$ twist 
fields $\mu$ simulate the branch point of a hyper-elliptic surface in case
of the $c=-2$ theory. Whenever all branch points have the same ramification
number, say $n$, all monodromies around these points can be diagonalized simultaneously.

As Knizhnik \cite{Knizhnik:1987xp} has shown, ghost systems on such
$\mathbb{Z}_n$-symmetric Riemann surfaces can be dealt with by putting them
on an $n$-fold sheeted covering of the complex plane where the branch points
are represented by suitable constructed vertex operators. However, these
vertex operators are twist fields, and thus may produce logarithmic
divergences in their operator product expansions. Furthermore, to yield 
a local theory, we have to take the tensor product of the theories on all 
covering sheets. 

The simplest such case is the hyper-elliptic one, since then automatically
all branch points are of order two. This hyper-elliptic case is special
since for the $c=-2$ theory, and only for this theory, one of the two
copies of the conformal field theory decouples completely.
This is a major difference of the $c=-2$ theory compared
to other ghost systems, namely that it is possible to eliminate the theory on
one of the two covering planes because after diagonalizing
the monodromies the vertex operators associated
to the branch cuts become trivial on one of the sheets.

Since this is a subtle point, we discuss it a bit more in detail:
The twist field $\mu$ for a
branch point on a hyper-elliptic surface for the $c=-2$ ghost system is
actually given by $\mu(z)=V_{-1/2}(z)\tensor V_{0}(z)$, where $V_q(z)$
denotes a vertex operator with charge $q$ with respect to the ghost current
$J=\mbox{:$bc$:}$ in a free field construction, 
and where we have indicated the composition of the
twist field out of the two copies of the CFT. The conformal weight is, with
$h(q)=\frac{1}{2}q(q+1)$, given by $h(-\frac{1}{2})+h(0)=-\frac{1}{8}$ as it
should be. The background charge at
infinity is for both copies $q_0=-1/2$ such that the total sum of all
charges in each copy must add up to $2q_0=-1$. Looking at the four-point
function mentioned above, we actually have to compute
\begin{eqnarray}\label{eq:twist4pt-again}
  \langle\mu(z_1)\mu(z_2)\mu(z_3)\mu(z_4)\rangle 
  &=& \langle V_{-1/2}(z_1)V_{-1/2}(z_2)V_{-1/2}(z_3)V_{-1/2}(z_4)\rangle \nonumber \\
  & & \times \, \langle V_{0}(z_1)V_{0}(z_2)V_{0}(z_3)V_{0}(z_4)\rangle\nonumber\\
  &=& \langle Q_{+1}V_{-1/2}(z_1)V_{-1/2}(z_2)V_{-1/2}(z_3)V_{-1/2}(z_4)\rangle \nonumber \nonumber\\
  & & \times \, \langle Q_{-1}V_{0}(z_1)V_{0}(z_2)V_{0}(z_3)V_{0}(z_4)\rangle\,,
\end{eqnarray}
where we have indicated the necessary screening charges in the last step.
Now, we can easily construct a screening current with charge $q=1$ since
$V_{1}(z)$ has conformal weight $h(q)=\frac{1}{2}q(q+1)=1$ as we expect.
Actually, $V_1(z)$ behaves essentially in the same way as the screening current,
since $J(z){\rm d}z = \mbox{:$bc$:}(z){\rm d}z$ transforms exactly like a
one-differential. 
Thus $Q_{+1} = \oint{\rm d}zV_{1}(z)$. This factor of the four-point function
yields then precisely the integral representation of the hyper-geometric
function appearing in (\ref{eq:twist4pt}). The second factor of the four-point
function is more tricky, since the field $V_{-1}$ has conformal weight $h=0$,
thus cannot serve as screening current. However, a screening current with the
correct properties can be constructed in the form $Q_{-1}=
\oint{\rm d}z\oint{\rm d}z'V_{1}(z)V_{-2}(z')$, since $V_{-2}$ also has
conformal weight $h=1$. When inserting these two screening charges, one has
to be careful with the choice of the contour for the integration. It turns
out that the net result in the presence of nothing but four identity fields
$V_0(z_i)$, $i=1,\ldots,4$, simply is the operator 
$\mbox{$:\phi V_{-1}:(0)$}$, where
$\phi(z)$ is the free field used in the bosonization. Thus, we end up with
the insertion of the {\em logarithmic\/} partner $\tilde{\mathbb{I}}(0)$ of the
identity such that the second factor of (\ref{eq:twist4pt-again}) does not
vanish identically, but yields simply a constant. 
Taken all together, we arrive at (\ref{eq:twist4pt}).

Repeating this computation for the $c=-26$ ghost system is a bit more
involved. The twist fields for the hyper-elliptic case have now the
composition $\mu(z)=V_{-1/2}(z)\tensor V_{-1}(z)$, such that the second
factor is not merely the identity operator. The conformal weights are
now given by $h(q)=\frac{1}{2}q(q+3)$ and the background charge at
infinity is now $-3/2$. The twist field has therefore conformal weight
$h_{\mu}=h(-\frac{1}{2})+h(-1)=-5/8+(-1)=-13/8$. Thus, we have to satisfy
\begin{eqnarray}\label{eq:twist4pt-c26}
  \langle\mu(z_1)\mu(z_2)\mu(z_3)\mu(z_4)\rangle 
  &=& \langle V_{-1/2}(z_1)V_{-1/2}(z_2)V_{-1/2}(z_3)V_{-1/2}(z_4)\rangle \nonumber\\
  & & \times \, \langle V_{-1}(z_1)V_{-1}(z_2)V_{-1}(z_3)V_{-1}(z_4)\rangle\nonumber\\
  &=& \langle Q_{-1}V_{-1/2}(z_1)V_{-1/2}(z_2)V_{-1/2}(z_3)V_{-1/2}(z_4)\rangle \nonumber \\
  & & \times \, \langle Q_{+1}V_{-1}(z_1)V_{-1}(z_2)V_{-1}(z_3)V_{-1}(z_4)\rangle\,,
\end{eqnarray}
where we have again indicated the necessary screenings. Here, the second
factor is easier, since the screening charge $Q_{+1}$ can always be taken as the
contour integration of the ghost current $J_{+1}(z)\equiv J(z)=
\mbox{:$bc$:}(z)$, 
since it transforms by construction as a one-differential. Moreover, all 
charges $q$ are always defined with respect to this ghost current. This is true
independent of the value of the spin $\lambda$ of the ghost system considered.
Thus, the screening charge $Q_{+1}$ is always easy to construct. 

For the first factor, we have to use a modified version of the screening
current, since the current $\tilde{J}(z)=\mbox{:$V_1V_{-2}$:}(z)$, although
it has the correct conformal weight $h=1$ and is a local chiral field,
does not yield the correct charge. It is merely an alternative representation
of the screening current. Instead, we might use
$J_{-1}(z)=\oint{\rm d}z'V_{1}(z)V_{-2}(z')=
\oint{\rm d}z'(z-z')^{-2}V_{-1}(z')$. This current has the correct
charge, but the wrong conformal weight $h=0$. We arrive thus at a similar
situation as with the second factor in the $c=-2$ case, namely where
the effect of screening is the insertion of a non-trivial $h=0$ field.

However, it is possible to construct a correct screening for the first
factor by making use of the non-trivial $h=5$ field of charge $q=2$, which
is part of the extended chiral symmetry algebra of the $c=-26$ ghost system.
The correct screening charge reads then
\begin{equation}
  Q_{-1} = 
  \oint{\rm d}u_1\oint{\rm d}u_2\oint{\rm d}u_3
  V_{-1}(u_1)V_{-2}(u_2)V_2(u_3)\,.
\end{equation}
The integrand has total conformal weight $h=(5)+(-1)+(-1)=3$,
which after three integrations yields a conformally invariant object. 

The lengthy discussion shows the following: Evaluating the four-point
function of four $\mathbb{Z}_2$ twist fields in the $c=-26$ case as in
(\ref{eq:twist4pt-c26}) yields an expression which will exhibit
logarithmic singularities just as in the $c=-2$ case. 
Indeed, the second
factor in the $c=-26$ case is again related to an integral representation
of a hyper-geometric system, ${}_2F_1(1,0;0;x)={}_1F_0(1;x)=\int_{x_0}^x{\rm d}u
(1-u)^{-2}$. The first factor,
however, is much more complicated since it involves a three-fold integration 
\begin{equation}
  \oint{\rm d}u_1\oint{\rm d}u_2\oint{\rm d}u_3
  \frac{(u_1-u_2)^2}{(u_1-u_3)^2(u_2-u_3)^4}\prod_{i=1}^4
  \frac{(z_i-u_1)^{1/2}(z_i-u_2)^1}{(z_i-u_3)^1}\,.
\end{equation}
After bringing the four-point function (\ref{eq:twist4pt-c26}) into
standard form with $z_1=\infty$, $z_2=1$, $z_3=x$, $z_4=0$ with $x$
the crossing ratio, one of the three integrations can be performed and yields
a Lauricella system of $D$-type (see for example \cite{Exton}), 
which is a generalized hyper-geometric system of several variables:
\begin{eqnarray}
  & & \langle V_{-1/2}(\infty)V_{-1/2}(1)V_{-1/2}(x)V_{-1/2}(0)\rangle\ =\
  \oint{\rm d}u_2\oint{\rm d}u_3(u_2-u_3)^{-4}\phantom{mm}\\
  & & \phantom{mmmm}\times F_D^{(3)}(\frac{3}{2},-\frac{1}{2},-2,2;3;x,u_2,u_3)
  \frac{u_3(1-u_3)(x-u_3)}{u_2(1-u_2)(x-u_2)}\,.\nonumber
\end{eqnarray}
The system $F_D^{(3)}$ has several solutions depending on the choice of the
integration contour, some of them exhibiting logarithms when expanded around
$x=0$. This is similar to the ordinary hyper-geometric case where a 
logarithmic solution appears whenever $c$ in ${}_2F_1(a,b;c;x)$ is an 
integer. In fact, $F_D^{(3)}(\frac{3}{2},-\frac{1}{2},-2,2;3;x,u,u)=
{}_2F_1(\frac{3}{2},-\frac{1}{2};3;x)$, which is a hyper-geometric system 
with the two expansions
\begin{eqnarray}
  y_1 &=& \phantom{\log(x)}\sum_n
          \frac{(\frac{3}{2})_n(-\frac{1}{2})_n}{(3)_n(1)_n}x^n\,,\\
  y_2 &=& \log(x)\sum_n
          \frac{(\frac{3}{2})_n(-\frac{1}{2})_n}{(3)_n(1)_n}x^n +
	  \sum_n\frac{\partial}{\partial\epsilon}\left(\frac{
	    (\frac{3}{2}+\epsilon)_n(-\frac{1}{2}+\epsilon)_n}{
	    (3+\epsilon)_n(1+\epsilon)_n}\right)_{\epsilon=0}x^n\ \ 
\end{eqnarray}
around $x=0$.
The full computation of this four-point functions is beyond the
scope of this paper.

We note once more that 
looking at $n$-point functions of twist fields reveals whether we
should expect logarithmic operators and thus indecomposable structures in
our CFT or not. The logarithmic operators get exchanged in the internal
channels of the $n$-point functions of twist fields due to degeneracies in the
moduli space of the considered Riemann surface, if branch points run into 
each other. The present case, $c=-26$, clearly shows all signs to be
a logarithmic CFT.


This discussion motivates, however, that our logarithmic deformation of the
ghost system is related to the above mentioned situation on nontrivial
Riemann surfaces. For the sake of simplicity, we concentrate again on the 
hyper-elliptic case. Doing so, we now have two sets of 
modes ($\theta^p_n, \btheta^p_n, \xi^p_i, \bxi^p_i, n \in \mathbb{Z}, p=1,2$) 
and also two sets of deformation parameters: $A_1, A_2, B_1, B_2$.

The easiest unification of both theories is given by defining
the modes of the unified theory in the following way:
\begin{align}
  L_n^{\mathrm{tot}} := L_n^{\mathrm{log},1} + L_n^{\mathrm{log},2} \quad ,
\end{align}
which means that we indeed take simply the tensor product of the two
isomorphic conformal field theories.
However this alone does not lead to a proper theory, since the
new modes $L_n^{\mathrm{tot}}$ do not satisfy the Virasoro algebra. To achieve 
the latter we have to identify the new modes, $\bxi_i, \theta_i$, on the one 
plane 
with the $\xi_i$, $\btheta_i$ modes on the other covering plane, by demanding
\begin{eqnarray}
  \theta_i^1  & \sim & \btheta_i^2 \\
  \bxi_i^1  & \sim & \xi_i^2 \quad .
\end{eqnarray}
and analogously for $\theta_i^2$ and $\bxi_i^2$.  
Using up two more degrees of freedom by setting
\begin{eqnarray}
  A := A_1 & = & - A_2 \\
  B := B_1 & = & - B_2
\end{eqnarray}
we get
\begin{align}
  \commu{L_n^{\mathrm{log},1}}{L_m^{\mathrm{log},2}} = 0 \quad , \label{eq:commu_2_planes}
\end{align}
and therefore $L_n^{\mathrm{tot}}$ now not only fulfills the Virasoro algebra
with total central charge $2 \cdot (-26)=-52$, but also acts correctly on 
the full space of states. It is worth mentioning that our construction 
automatically and naturally forces us to consider the (deformed) ghost
system conformal field theory on a nontrivial Riemann surface. 
Moreover, we also have to slightly alter Knizhnik's prescription of
constructing the full conformal field theory. A consistent Virasoro algebra
with the correct action on the Hilbert space
can only be obtained, if the two copies are not simply added, but only
if the \zeromodes\ of the two conformal field theories are intermixed.
This, in essence, encodes that the action of the monodromies cannot be
fully diagonalized, leading to indecomposable structures in the conformal
field theory. It is very interesting that for $c=-26$, and presumably for
any other ghost system with $\lambda\neq 1$, the deformation of the
Virasoro algebra inevitably leads us to consider such tensor products of
these ghost systems, which do not factorize completely. As mentioned above,
the $c=-2$ case appears now as particularly simple, since here the
factorization of the full theory in two copies still almost holds.\footnote{
Of course, one should in principle also identify the additional \zeromode\ 
for the deformation with the \zeromode\ of the other copy of the conformal
field theory for the other sheet.} Thus, our enlarged theory has a nice
and natural geometrical interpretation.

Applying the same steps to our second solution 
(\ref{eq:second_sol_start})-(\ref{eq:second_sol_end})
gives:
\begin{eqnarray}
  A' := {A'}_1 & = & - {A'}_2 \\
  C' := {C'}_1 & = & - {C'}_2 \\
  B' := {B'}_1 & = & {B'}_2 = 0 \quad .
\end{eqnarray}
This means that the second solution is already included
in the first one ($B=0$) and in particular the condition (\ref{eq:commu_2_planes}) 
enforces the elimination of the terms containing $\xi$ modes.
Therefore, it is sufficient
to investigate the first solution though we bear in mind
that $B=0$ might be an interesting choice.

Retranslating the system to the familiar \bcsystem\ 
using the choice above leads to
\begin{eqnarray}
  R_{-2}^{\mathrm{tot}} & = & - \frac 12 A b^1_{-3} b^2_1 - 2 B b^1_{-2} b^2_0
         - 3 A b^1_{-1} b^2_{-1} + \frac 12 A b^2_{-3} b^1_1
         + 2 B b^2_{-2} b^1_0 \label{eq:R_-2} \\
  R_{-1}^{\mathrm{tot}} & = & -(A+B) b^1_{-1} b^2_0 - (A+B) b^1_0 b^2_{-1} \\
  R_{0}^{\mathrm{tot}} & = & 0 \\
  R_{1}^{\mathrm{tot}} & = & (A+B) b^1_0 b^2_1 + (A+B) b^1_1 b^2_0 \\
  R_{2}^{\mathrm{tot}} & = & 3 A b^1_1 b^2_1 + 2 B b^1_2 b^2_0
         + \frac 12 A b^1_3 b^2_{-1} + 2 B b^1_0 b^2_2 + \frac 12 A b^1_{-1} b^2_3 \, .
\end{eqnarray}
Therefore, our theory is diagonal with respect to $L_0^{\mathrm{tot}}$
for arbitrary $A$ and $B$. Off-diagonal contributions appear 
in all different modes for almost all nontrivial choices of $A$ and $B$. 
The only nontrivial exception is $A=-B$ which eliminates 
all off-diagonal elements for $L_{-1}^{\mathrm{tot}}$ and $L_1^{\mathrm{tot}}$
thus leading to a theory which is as ``little'' as possible logarithmic, in
the sense that the $\mathrm{SL}(2,\mathbb{C})$ global conformal group
is not deformed at all. In particular the second solution 
(\ref{eq:second_sol_start})-(\ref{eq:second_sol_end})
which narrowed down to the first one with $B=0$ 
comes for all nontrivial choices always with a deformation term.
Note that there is no physical reason forcing 
this choice.\footnote{If we want to keep the relations we already 
know from other LCFTs, then $A=-B$ is mandatory. 
In the so far known LCFTs, the Virasoro modes can be written in 
the form $L_n = z^n \left(z\partial_i + (n+1)(h_i + \delta_{h_i})\right)$,
see e.\,g.\ \cite{Flohr:2000mc}, such that $L_{-1}$ has no
off-diagonal contribution, which one might expect for the 
generator of translations. This differs from the case considered
here, where $L_0$ has no off-diagonal term.
It follows then from the Virasoro algebra that $L_{-1}$
having no logarithmic contribution implies the same for
$L_1$ and vice versa. The choice $A=-B$ reproduces this behavior.} For 
any nontrivial choice of $A$ and $B$ it is inevitable that deformations 
of higher modes $|n| \ge 2$ occur. 

The next question is, what the highest weight states for our
enlarged theory are, since these correspond to the primary fields.
Leaving aside twisted sectors of the theory,
we found the following highest weight states for $h = -2, -1, 0$ (note that there are no
such states for $h = 1, 2$ and that all states for $h=-2$ are highest weight states).
\begin{align}
  h = -2: \quad & c^{1}_{1} c^{2}_{1} \vacuum,
            c^{1}_{0} c^{1}_{1} c^{2}_{1}\vacuum, c^{1}_{1} c^{2}_{0} c^{2}_{1}\vacuum, 
            c^{1}_{0} c^{1}_{1} c^{2}_{0} c^{2}_{1}\vacuum \nonumber \displaybreak[0] 
    \\
         & \nonumber \displaybreak[0] \\
  h = -1: \quad & c^{2}_{1} \vacuum, c^{1}_{1}\vacuum, \nonumber \\
          & c^{1}_{0} c^{1}_{1} \vacuum, \left( c^{1}_{0} c^{2}_{1} - c^{1}_{1} c^{2}_{0}\right) \vacuum, 
            c^{2}_{0} c^{2}_{1}\vacuum, \nonumber\\
          & \left(c^{1}_{0} c^{1}_{1} c^{2}_{0} -2 c^{1}_{-1} c^{1}_{1} c^{2}_{1}\right)\vacuum, 
            \left(c^{1}_{0} c^{2}_{0} c^{2}_{1} -2 c^{1}_{1} c^{2}_{-1} c^{2}_{1}\right)\vacuum \nonumber \\
          & c^{1}_{-1} c^{1}_{0} c^{1}_{1} c^{2}_{1}\vacuum, 
            \left(c^{1}_{0} c^{1}_{1} c^{2}_{-1} c^{2}_{1} - 
              c^{1}_{-1} c^{1}_{1} c^{2}_{0}c^{2}_{1}\right)\vacuum,
            c^{1}_{1} c^{2}_{-1} c^{2}_{0} c^{2}_{1}\vacuum, \nonumber \\
          & - ( A + B ) c^{1}_{-1} c^{1}_{1} c^{2}_{1}\vacuum
            +  c^{1}_{-1} c^{1}_{0} c^{1}_{1} c^{2}_{0} c^{2}_{1}\vacuum, \nonumber \\
          & ( A + B ) c^{1}_{1} c^{2}_{-1} c^{2}_{1} \vacuum
            + c^{1}_{0} c^{1}_{1} c^{2}_{-1} c^{2}_{0} c^{2}_{1} \vacuum \nonumber \displaybreak[0] \\
          & \nonumber \displaybreak[0] \\
  h = 0: \quad  & \vacuum, \nonumber \\
          & c^{1}_{-1} c^{1}_{0} c^{1}_{1} \vacuum, c^{2}_{-1} c^{2}_{0} c^{2}_{1} \vacuum, \nonumber \\
          & \left( c^{1}_{-1} c^{1}_{1} c^{2}_{0}  - c^{1}_{-1} c^{1}_{0} c^{2}_{1}
            - c^{1}_{0} c^{1}_{1} c^{2}_{-1} \right) \vacuum, \nonumber \\
          & \left( c^{1}_{-1} c^{2}_{0} c^{2}_{1} + c^{1}_{1} c^{2}_{-1} c^{2}_{0}
            - c^{1}_{0} c^{2}_{-1} c^{2}_{1} \right) \vacuum, \nonumber \\
          & ( \frac{A^2}{4}  c^{1}_{0} c^{2}_{0} + \frac 12 A B c^{1}_{-1} c^{2}_{1}  
            + \frac 12 A B c^{1}_{1} c^{2}_{-1} + B c^{1}_{-1} c^{1}_{1} c^{2}_{-1} c^{2}_{1} \nonumber\\
          & + \frac 12 A c^{1}_{-1} c^{1}_{0} c^{2}_{0} c^{2}_{1}
            + \frac 12 A c^{1}_{0} c^{1}_{1} c^{2}_{-1} c^{2}_{0} 
            + c^{1}_{-1} c^{1}_{0} c^{1}_{1} c^{2}_{-1} c^{2}_{0} c^{2}_{1} ) \vacuum   \label{eq:tot_hws0}
\end{align}
As we noted above $L^{\textrm{tot}}_0$ is---in contrast to the $c=-2$ theory---diagonal. 
An operator for $c=-26$ which has similar properties as $L_0$ for $c=-2$
is $L^{\textrm{tot}}_{-2}$. Indeed, applying this operator 
generates off-diagonal terms as the following example shows
\begin{align}
L^{\textrm{tot}}_{-2} \ket{ c^1_1 c^2_1 } = 
         \left( c^1_1 c^2_{-1} + c^1_{-1} c^2_1 \right) \vacuum
         - 2 \left( b^2_{-2} c^2_0 + b^1_{-2} c^1_0 \right) \ket{c^1_1 c^2_1} + 3A \vacuum \, .
\end{align}
Applying $L^{\textrm{tot}}_{-2}$ a second time leads to further off-diagonal terms
\begin{align}
  \left( L^{\textrm{tot}}_{-2} \right)^2 = &
    - A ( 12 b^1_{-2}c^1_0 + \frac 32 b^1_{-3}c^1_1 
    + 12 b^2_{-2}c^2_0 + \frac 32 b^2_{-3}c^2_1)\vacuum \nonumber \\
  & + 8 B ( b^1_{-2} c^1_1 b^2_{-2} c^2_1)\vacuum \, .
\end{align}
While $A=-B$ in general makes the theory easier (by eliminating
logarithmic contributions) this does not reduce the number
of terms in this case. 

If we multiply the deformation term with $q$ ($A \rightarrow qA$, $B \rightarrow qB$)
then it is interesting to note that the power in $q$
does not go beyond a certain threshold if we consider
$(L^{\textrm{tot}}_{-2})^m \ket{\mathrm{state}}, \, m \in \mathbb{N}$. 
The reason for a threshold can be derived from
the structure of the extension $R^{\mathrm{tot}}_{n}$
and the states: 
each $R^{\mathrm{tot}}_{n}$ contains 
at least one annihilator $b^p_i \, (p=1,2,\, i=-1,0,1)$ while 
the states are words in the conjugated modes, the creators,
$c^p_i \, (p=1,2, i=-1,0,1)$ applied to the \invacuum.
By applying $R^{\mathrm{tot}}_{n}$ the number of $c$ modes is reduced 
by one or two (or the term is eliminated) and most importantly there is no term 
in $L^{\mathrm{tot}}_{n}$ which increases the number of $c$ modes again.
Therefore, the maximum power in $q$ which theoretically can occur is $6$.

Our $c=-52$ theory comes, though logarithmic, with a 
non-logarithmic $L^{\mathrm{tot}}_{0}$ which 
is a major difference to all LCFTs we know up to now.  
Because of $L^{\mathrm{tot}}_{0}$ being trivial we obviously get no 
\jordancell\ or a \jordanrank\ in the traditional sense. 
Nevertheless we have some properties which are the same in both types of LCFTs,
the ones with and without logarithmic $L^{\mathrm{LCFT}}_{0}$. Remember that 
applying $L^{\mathrm{LCFT}}_0$ on a highest weight state $\ket{h,k}$ 
leads to an extra term $\ket{h,k-1}$ for $k > 0$. Therefore, marking the 
logarithmic extension term with a $q$ leads to
\begin{equation}
  (L^{\mathrm{log}}_0)^m \ket{h,k}  = q^k \ket{h,0} + q^{k-1}(\ldots) + 
                                      \ldots + h^m\ket{h,k} \quad , m > k
\end{equation}
where $k = 0, \ldots, \mathrm{jrk} \, (L^{\mathrm{log}}_0) - 1$
and $\mathrm{jrk}$ denotes the rank of the \jordanmatrix.
This means that the \jordanrank\ can be found by applying 
$(L^{\mathrm{LCFT}}_0)^m$ for all $m \in \mathbb{N}$ on
all $\ket{h} \in \mathrm{HWS}$ ($\mathrm{HWS}$ denotes the set of 
all highest weight states). The highest occurring power in $q$ plus $1$ defines
the rank of the \jordancell. This motivates the following definition:
writing $L^{\mathrm{LCFT}}_n = L^{\mathrm{CFT}}_n + q  R_n$
where $R_n$ is the deformation term
\begin{equation}
  \mathrm{jrk} \, (L^{\mathrm{log}}_n) 
    := \max \left\{ k = \mathrm{deg}_q\left( (L^{\mathrm{log}}_0)^m \ket{h} \right) 
                    : \ket{h}\in\mathrm{HWS}, m\in\mathbb{N}  \right\} \, .
\end{equation}
where $(L^{\mathrm{log}}_0)^m \ket{h}$ is to be understood as
a polynomial in $q$ after evaluation.
The logarithmic behavior of this theory becomes (for $A = -B$)
manifest in $L^{\mathrm{tot}}_{-2}$. The \jordanrank\ in the 
above defined sense of $L^{\mathrm{tot}}_{-2}$
can easily be found by examining (\ref{eq:R_-2}): each term contains at least 
one of the modes $b^p_i \, (p=1,2, i=-1,0,1)$. The remaining $b$ modes are of
no interest since these are creators and the zero modes
are mutually distinct in each of the terms. Therefore, the only states
we are interested in are words in the letters 
$c^2_{-1}, c^2_0, c^1_{1} c^2_{1}, c^1_{-1}, c^1_0$.
Looking at the highest weight states of conformal weight $h=0$ 
in eq.\ (\ref{eq:tot_hws0}) shows that one highest weight state really contains
a state which consists of all the above letters implying an upper bound
\begin{align}
  \mathrm{jrk} \, (L^{\mathrm{tot}}_{-2}) = 5
\end{align}
up to accidental cancellations. Tedious and lengthy calculations
reveal that the upper bound is satisfied. We note for completeness 
that for $B=0, A\ne0$ the \jordanrank\ is $\mathrm{jrk} \, (L^{\mathrm{tot}}_{-2}) = 3$.

\section{Summary \& Conclusion}
The well-known \bcsystem\ with central charge $c=-26$
can be enlarged to a logarithmic CFT. In some aspects
the transition is similar to the $c=-2$ case,
in others it is completely different: the energy momentum tensor
cannot be built by combining derivatives of the 
generalized symplectic fermion fields and we also have to consider 
that it is not correct to neglect one half of the theory 
if we investigate it on hyper-elliptic Riemann surfaces.
On the contrary, enlarging the $c=-26$ ghost system to a logarithmic
theory makes it necessary to consider this theory on nontrivial
Riemann surfaces. This is natural and consistent with our understanding
of the geometrical origin of logarithmic fields from operator product
expansions of twist fields which simulate branch points. On the other hand,
it is surprising in so far as it is possible to consider the logarithmic
extension of the better known $c=-2$ theory without the need of putting it on 
higher genus Riemann surfaces. As we have seen, this is impossible for
$c=-26$. Due to the particular structure of the vertex operators which
represent the branch points, we conjecture that logarithmic extensions
of other ghost systems with $\lambda\neq 1,2$ are only possible when
considered on $\mathbb{Z}_n$-symmetric Riemann surfaces.

We are confident that the presented construction 
scheme works not only for $c=-26$ but for all \bcghostsystems.
The deformation term we used in order to obtain the new 
energy-momentum tensor is slightly more general than the 
deformation term discussed in the paper by
Fjelstad, Fuchs \emph{et al.} \cite{Fjelstad:2002ei}, but is naturally linked to the \zeromode\ 
structure of the ghost systems. Thus, we expect that the spin
$(\lambda,1-\lambda)$ ghost system has generically $2\lambda+1$ deformation
directions which presumably get restricted due to hermiticity conditions
and consistency requirements for the action of the deformed Virasoro algebra
of the full theory on the Hilbert space of states.

The structure of the logarithmic $c=-26$ theory is very
different from what one might have expected in analogy to the $c=-2$ case: 
$L^\mathrm{tot}_0$ 
is not logarithmic at all. This is a completely new
property of a LCFT. Furthermore the special choice
of the deformation parameter $A=-B$ (see eq.\ (\ref{eq:Tlog_b_theta}))
leads to a theory where the whole global
conformal group is non-logarithmic. This special
property is not yet completely investigated.
The logarithmic character of the theory 
becomes manifest in $L^\mathrm{tot}_{-2}$
which shows similar indecomposable properties 
as $L_0$ in a standard LCFT.
A generalization of the definition of the \jordanrank\ has been 
given which we used to find that the \jordanrank\ of $L^\mathrm{tot}_{-2}$ is $5$ 
for all nontrivial choices of $A$ and $B$. This should help in identifying the
proper generalization of ``logarithmic partners'' to primary fields,
which is left for future work.

\section*{Acknowledgment} The research of M.\,F.\ is supported by
the European Union network HPRN-CT-2002-00325 and the research of M.\,F.\ and 
M.\,K.\ is supported by the string theory network
(SPP no.\ 1096), Fl 259/2-1, of the Deutsche Forschungsgemeinschaft.
We thank Sebastian Uhlmann and the participants of the workshop 
{\em Non-Unitary \&
Logarithmic Conformal Field Theory}, which was held 10--14 June 2002
at the Institute des Hautes \'Etudes Scientifiques, Paris, France, for
numerous discussions.

\bibliographystyle{utcaps}
\bibliography{hep-th}

\end{document}